\documentclass[12pt]{article}
\usepackage{epsfig,amsmath,amsfonts,verbatim}
\usepackage{lmodern}
\usepackage[T1]{fontenc}
\usepackage{mathtext,marvosym,textcomp}
\usepackage{indentfirst}
\usepackage{epsfig,amsmath,amsfonts,amssymb}
\usepackage{mathtools}
\usepackage[usenames,dvipsnames,svgnames,table]{xcolor}
\usepackage{tikz}
\usetikzlibrary{arrows}
\usetikzlibrary{shapes.misc}
\usetikzlibrary{shapes,positioning,matrix,arrows}
\tikzstyle{every picture}+=[remember picture]
\tikzstyle{na} = [baseline=-.5ex]
\tikzstyle{format} = [rectangle,
                      thick,
                      minimum size=2cm,
                      draw=red!00!black!50,
                      top color=white,
                      bottom color=red!00!black!00,
                      text centered,
                      font=\fontsize{10}{10}\selectfont,
                      on grid]
\tikzstyle{format1} = [rectangle,
                      thick,
                      dashed,
                      minimum size=2cm,
                      draw=red!00!black!50,
                      top color=white,
                      bottom color=red!00!black!00,                     
                      text centered,
                      font=\fontsize{10}{10}\selectfont,
                      on grid]
\tikzstyle{format2} = [font=\fontsize{10}{10}\selectfont,
                      on grid]
\tikzset{cross/.style={cross out, draw=black, minimum size=2*(#1-\pgflinewidth), inner sep=0pt, outer sep=0pt},
cross/.default={5pt}}
 \usepackage[ normalem]{ulem}
 \usepackage{tocloft}

\usepackage[debug,pageanchor=false]{hyperref}
\hypersetup{colorlinks=true
,urlcolor=blue
,anchorcolor=blue
,citecolor=blue
,filecolor=blue
,linkcolor=blue
,menucolor=blue
,pagecolor=blue
,linktocpage=true
,pdfproducer=medialab}

\numberwithin{equation}{section}

\def\a{\alpha} \def\b{\beta} \def\g{\gamma} \def\d{\delta} 
   
    \def\m{\mu}
  \def\p{\pi}  \def\r{\rho}
 \def\s{\sigma}

  \def\P{\Pi}

\def\fr{\frac}  \def\dt{\partial}

\def\ph{\phantom}
\def\mc{\mathcal}

\newcommand\bqa {\begin{eqnarray}}
\newcommand\eqa {\end{eqnarray}}

\newcommand{\bear}{\begin{array}}
\newcommand{\enar}{\end{array}}

\def\beq{\begin{equation}}
\def\eeq{\end{equation}}
\def\bea{\begin{eqnarray}}
\def\eea{\end{eqnarray}}

\begin{document}

\begin{titlepage}

\ph{}

\begin{center}
   \baselineskip=12pt
   \centerline{ {\Large \bf 
   Integrability properties of Motzkin polynomials}}
   \vskip 2cm
    {\large {\bf Ilmar Gahramanov$^{\star\times\$}$\footnote{\tt ilmar.gahramanov@aei.mpg.de} and  Edvard T. Musaev$^{\star\dagger}$\footnote{\tt edvard.musaev@aei.mpg.de}}}
       \vskip .6cm
             \begin{small}
                          {\it
                          $^\star$Max-Planck-Institut f\"ur Gravitationsphysik (Albert-Einstein-Institut)\\
                          Am M\"uhlenberg 1, DE-14476 Potsdam, Germany\\[0.5cm] 
                          $^\times$Institute of Radiation Problems ANAS,\\ 
                          B.Vahabzade 9, AZ1143 Baku, Azerbaijan;\\ Department of Mathematics, Khazar University, \\ Mehseti St. 41, AZ1096, Baku, Azerbaijan \\[0.5cm]
                          $^\$$Department of Physics, Mimar Sinan Fine Arts University,\\ Bomonti 34380, Istanbul, Turkey\\[0.5cm]
                           $^\dagger$Kazan Federal University, Institute of Physics\\
                          General Relativity Department\\
                          Kremlevskaya 16a, 420111, Kazan, Russia                         
                          } \\ 
\end{small}
\end{center}

\vspace{1cm}
\begin{center} 
\textbf{Abstract}
\end{center} 
\begin{quote}
We consider a Hamiltonian system which has its origin in a generalization of exact renormalization group flow of matrix scalar field theory and describes a non-linear generalization of the shock-wave equation that is known to be integrable. Analyzing conserved currents of the system the letter shows, that these follow a nice pattern governed by coefficients of Motzkin polynomials, where each integral of motion corresponds to a path on unit lattice.
\end{quote} 
\vfill

\setcounter{footnote}{0}

\end{titlepage}

\clearpage
\setcounter{page}{2}

\tableofcontents

\section{Introduction}

Exact renormalization group governed by the Polchinski equation allows to investigate dynamics of operators in a field theory under change of scale \cite{Polchinski:1983gv,Bervillier:2004mf}. In the work \cite{Akhmedov:2010mz} it has been shown that for the simplest case, matrix scalar field theory, these RG flow equations can be written in Hamiltonian form for some specific choice of operators built from fundamental fields. The corresponding Hamiltonian was shown to be of the following form
\begin{equation} 
H=\int_0^{2\p} d\s\,\P^2J',
\end{equation}
that gives the so-called Burgers-Hopf equations 
\begin{equation}
\label{BH}
\dt_T\r= \r \dt_s \r, \quad \r= \fr{\dt_T{J}}{\dt_\s J},
\end{equation}
where $T$ is related to the physical energy scale at which observations are performed, and plays the role of time for the system. This equation describes shock waves and is known to be integrable. 

However, on the field theory side the above discussion is only valid in the IR limit, where a significant amount of information is allowed to be dropped off, hence leading to RG equations written in such ultra-local form. Attempts to go beyond this approximation and to obtain a more general Hamiltonian has been made in \cite{Akhmedov:2010sw}, which however were not completely successful. Despite the technical difficulties which so far has not allowed to arrive at an ultra-local Hamiltonian and which are related to cut-off function defining the physical energy scale, some speculations lead to the Hamiltonian given by
\begin{equation} \label{Ham}
H=\int_0^{2\p} d\s\,\left[\P^2J'+\P{J'}^2\right],
\end{equation}
whose limit $J \ll \P$ gives the previous system. Since the currents $J$ should vanish at low energies, this is in agreement with the IR nature of the previous Hamiltonian.For example, one of the ways to properly arrive to the above Hamiltonian would be to redefine fundamental fields of the initial system in order to hide the cut-off function inside. This in turn leads to other technical difficulties and hence will not be considered in the present letter. 

Instead, we start from the Hamiltonian \eqref{Ham} as it is and investigate it's integrability properties motivated by the above discussion. In the spot of our interest here are integrals of motions, equivalently conserved currents, which can be explicitly constructed and parametrized by an integer number. All these stay in involution with respect to the naturally defined Poisson brackets, that might be a good sign of (classical) integrability of the system. Another argument in favor of this conjecture is that taking the limit $J \ll \P$ and hence dropping the second term in the Hamiltonian one is returns back to the Burgers-Hopf shock wave equations that is known to be integrable. However, after trying to apply known to us integrability criteria, such as constructing a Lax pair and performing the Painlev\'e test, we were not able to show this explicitly so far and hence leave this task for a further work. 

The main result of the letter is observation of the so-called Motzkin numbers in this Hamiltonian system, which usually appear  in description of unit paths (see sequence A055151 in \cite{Sloane} and references therein). We show that in the considered system these numbers appear as coefficients in the obtained integrals of motion thus rendering them precisely as Motzkin polynomials and allowing to rewrite them as a very nice expression in terms of hypergeometric ${}_2F_1$ functions. This might be an interesting result from the pure mathematical point of view. In the conclusion section we discuss possible extensions of the current result.

\section{Motzkin numbers}

In this section we give a short review describing Motzkin paths and related polynomials which appear in many contexts in the mathematical literature, e.g. \cite{oste,krat,flajolet} and physics, e.g. \cite{blythe,bravyi,francesco2010q}.

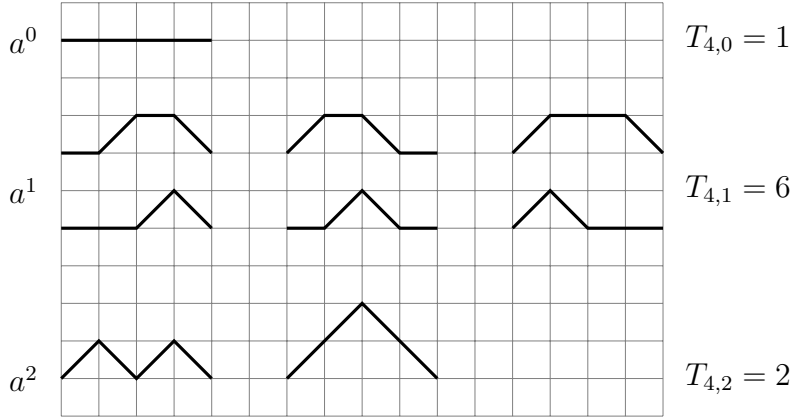
\begin{figure}[ht]
\centering
\begin{tikzpicture}[
                    ]
\draw[step=0.5, very thin, gray] (0.999,-5.5) grid (9,0);	
	
\newcommand\ya{0.5};
\draw (0.5,-\ya) node {$a^0$};
\draw[very thick] (1,-\ya) -- (3,-\ya);
\draw (10,-\ya) node {$T_{4,0}=1$};

\newcommand\yb{2};
	\draw (0.5,-\yb-0.5) node {$a^1$};
	\draw (10,-\yb-0.5) node {$T_{4,1}=6$};	
	
	\draw[very thick] (1,-\yb)--(1.5,-\yb)--(2,-\yb+0.5)--(2.5,-\yb+0.5)--(3,-\yb);
	\draw[very thick] (1,-\yb-1)--(1.5,-\yb-1)--(2,-\yb-1)--(2.5,-\yb-0.5)--(3,-\yb-1);

	\newcommand\xba{3};
	\draw[very thick] (1+\xba,-\yb)--(1.5+\xba,-\yb+0.5)--(2+\xba,-\yb+0.5)--(2.5+\xba,-\yb)--(3+\xba,-\yb);
	\draw[very thick] (1+\xba,-\yb-1)--(1.5+\xba,-\yb-1)--(2+\xba,-\yb-0.5)--(2.5+\xba,-\yb-1)--(3+\xba,-\yb-1);

	\newcommand\xcb{6}
	\draw[very thick] (1+\xcb,-\yb)--(1.5+\xcb,-\yb+0.5)--(2+\xcb,-\yb+0.5)--(2.5+\xcb,-\yb+0.5)--(3+\xcb,-\yb);
	\draw[very thick] (1+\xcb,-\yb-1)--(1.5+\xcb,-\yb-0.5)--(2+\xcb,-\yb-1)--(2.5+\xcb,-\yb-1)--(3+\xcb,-\yb-1);

\newcommand\yc{5};
	\draw (0.5, -\yc) node {$a^2$};
	\draw (10,-\yc) node {$T_{4,2}=2$};
		
	\draw[very thick] 
	(1,-\yc)--(1.5,-\yc+0.5)--(2,-\yc)--(2.5,-\yc+0.5)--(3,-\yc);     
	
	\newcommand\xca{3};
	\draw[very thick] 
	(1+\xca,-\yc)--(1.5+\xca,-\yc+0.5)--(2+\xca,-\yc+1)--
	(2.5+\xca,-\yc+0.5)--(3+\xca,-\yc);     	

\end{tikzpicture}
\caption{Motzkin paths of length 4 between the points (0,0) and (4,0). The powers of $a$ in the left column denote the number of up steps. The number of paths of length $n$ with $k$ up step is given by the Motzkin polynomial coefficient $T_{n,k}$.}
\label{Motz}
\end{figure}

First let us define a lattice path. A lattice path $L$ in $\mathbb{Z}^d$ of length $n$ is a sequence $v_0, \ldots, v_n \in \mathbb{Z}^d$ with corresponding steps $s_1, \ldots, s_n \in \mathbb{Z}^d$ defined by consecutive difference $s_i=v_i-v_{i-1}$. A Motzkin path of length $n$ is a lattice path on $\mathbb{N} \times \mathbb{N}$ consisting of up steps $(1, 1)$, down steps $(1, -1)$ and flat steps $(1,0)$. The number of Motzkin paths from $(0, 0)$ to $(n, 0)$ is given by the Motzkin number\footnote{Note that historically the Motzkin numbers appeared in a circle chording setting \cite{motzkin}.} (sequence A001006 in \cite{Sloane}) $m_n$ which can be written in the following form 
\begin{eqnarray}
m_n = \sum_{k=0}^{ [n/2]} {n \choose 2k} c_k \;,
\end{eqnarray}
where $c_k$ are the Catalan numbers (sequence A000108 in \cite{Sloane}) defined as
\begin{equation}
c_k = \frac{1}{k+1} {2k \choose k} \;.
\end{equation}
We are particularly interested in Motzkin polynomial associated to a Motzkin path. In order to define the Motzkin polynomial one needs to assign a weight keeping track of the number of up steps (or flat steps). Then one gets the following polynomial  with the corresponding coefficients (sequence A055151 in \cite{Sloane})
\begin{equation} \label{motpol}
m_n(a) \ = \ \sum^{[n/2]}_{k=0} T_{n,k} \; a^{k} 
\end{equation}
where $a$ stands for up steps. The coefficient $T_{n,k}$ is number of Motzkin paths of length n with k up steps. Note that the polynomial (\ref{motpol}) is also known as the Jacobi-Rogers polynomial \cite{flajolet}.

It might now be reasonable to expect the occurrence of other lattice path polynomials such that the Riordan polynomials, Dick polynomials etc. in the context of integrable models.

\section{Hamiltonian system for Motzkin numbers}

Consider a dynamical system defined by the following Hamiltonian
\begin{equation}
\label{H}
H=\int_{-\pi}^{\pi}d\s\, \left[\P^2J'+\P{J'}^2\right],
\end{equation}
that is motivated by the renormalization group procedure as discussed in the Introduction. Given the direct relation between integrability properties of this system and Motzkin paths and that the system does not look recognizable to the knowledge of the authors, we suggest to call it the Motzkin system\footnote{Note that there are papers \cite{doi:10.1063/1.4977829, chen2017quantum} on quantum spin chains where authors also use the terminology ``the Motzkin Hamiltonian'' for the specific nearest neighbor Hamiltonian \cite{bravyi}.}.

Let us start with equations of motion
\begin{equation}
 \label{4}
\begin{split}
 & \dot{J}=2\P J'+{J'}^2;\\
 & \dot{\P}=2\P\P'+2\P' J'+2\P J'',
\end{split}
\end{equation}
which can be easily solved with respect to $\P$ providing the following Lagrangian formulation of the theory
\begin{equation}
 \label{9}
 \mc{L}=\fr{\left(\dot{J}-{J'}^2\right)^2}{4J'}.
\end{equation}

The EOM for the field $J(t,\s)$ can be found by either varying the Lagrangian \eqref{9} with respect to $J(t,\s)$ or substituting an expression for $\P(t,\s)$ into the second equation in \eqref{4}:
\begin{equation}
 \label{10}
\begin{aligned}
&\dot{\r}-\fr12\dt_\s\r^2=5\dot{J}'-6J''J',\\
&\r:=\fr{\dot{J}}{J'}.
\end{aligned}
\end{equation}
One notices, that in the IR limit of the corresponding field theory the sources become infinitesimally small $J\to0$ , and hence the above equation drops to the Burgess-Hopf equation in agreement with \cite{Akhmedov:2010mz}.

However, in general the  above equation does not immediately correspond to one of the commonly known types of non--linear equations. On the other hand, if one is lucky to find the explicit Lax pair for these equations it would be possible either to compare the system to one of the known systems (e.g. KdV) or to prove that it is a new integrable system. However, so far we have not been able to find the corresponding Lax pair and we cannot claim if there exists one. In general the process of finding of the Lax pair is always some kind of an art.  

To move forward in analysis of the system, we present an infinite set of integrals of motion for the system in question, which appears to be a good argument in favors of integrability of the Hamiltonian flow. It is suggestive to consider the notion of integrability in the Liouville sence, that means having a maximal set of Poisson-commuting integrals of motion (i.e. function(al)s on the phase space whose Poisson brackets vanish) which are not trivial, i.e. not zero and not Casimir elements. Certainly one must be careful when applying the Liouville criterion to infinitely dimensional systems such as field theoretical equations. For this reason we present the required full set of integrals of motion refraining from the claim that the system is indeed completely integrable.

Hence, starting with some obvious integrals of motion of the type
\begin{equation}
 \label{5}
\begin{split}
 &I_1=\int_\s\left(\P+J'\right);\\
 &I_2=\int_\s\P J';\\
 &I_3=\int_\s \P J'(\P+J')=H,
\end{split}
\end{equation}
it is straightforward to see that there exists the following infinite tower of such constructions (see Appendix \ref{deriv1})
\begin{equation}
\label{6}
\begin{aligned}
 I_{n}&=\int_\s\sum_{k=1}^{n}\left(\P J'\right)^{k}\left(\P+J'\right)^{n-2k}t_{n,k}+\d_n^1 (\P+J'),\\
 t_{n,k}&=\fr{(n-2)!}{(n-2k)!k!(k-1)!}, \forall n>1\\
 t_{1,1}&=1.
 \end{aligned}
\end{equation}
It is worth mentioning here that one shouldn't be confused by the fact that we have discrete set of integrals of motion facing ``continuous'' variables $J(t,\s)$. Since $\s$ is compact one actually has a discrete spectrum of variables. 

By making use of the relation between $t_{n,k}$ and the Motzkin polynomial coefficients
\begin{equation}
t_{n,k}=T_{n-2,k-1}
\end{equation}
we can write
\begin{equation}
\begin{aligned}
I_n&=\int_\s\sum_{k=1}^n\left(\P J'\right)^{k}\left(\P+J'\right)^{n-2k}T_{n-2,k-1}+\left(\P+J'\right) \d_{n}^1.
\end{aligned}
\end{equation}

Interestingly, the sum in the first term above can be performed explicitly and the result for this term can be written in terms of hypergeometric function to give (note that $n\neq 1$)
\begin{equation}
\begin{aligned}
\bar{I}_n&=\int_\s \sum_{k=1}^n\left(\P J'\right)^k \left(\P+J'\right)^{n-2k}T_{n-2,k-1}\\
&=\int_\s \left(\P J'\right) \left(\P+J'\right)^{n-2}{}_2F_1\Big[1-\fr n2, \fr32-\fr n2;2,\fr{4\left(\P J'\right)}{\left(\P+J'\right)^2}\Big].
\end{aligned}
\end{equation}
This expression can be further simplified by making use of the following quadratic relation
\begin{equation}
\begin{aligned}
{}_2F_1\Big[\fr a2,\fr a2+\fr12;1+a-b,\fr{4z}{(1+z)^2}\Big]&=(1+z)^a{}_2F_1[a,b;1+a-b,z]
\end{aligned}
\end{equation}
to obtain
\begin{equation}
\bar{I}_n=\int_\s \P J'^{n-1}{}_2F_1\bigg[2-n,1-n;2,\fr{\P}{J'}\bigg].
\end{equation}
The conserved quantities $I_k$ can be shown to be in involution, i.e they commute with respect to the standard Poisson bracket
\begin{equation}
\begin{aligned}
\{f,g\}&=\fr{\dt f}{\dt J}\fr{\dt g}{\dt \P}-\fr{\dt f}{\dt \P}\fr{\dt g}{\dt J},\\
\{I_m,I_n\}&=0.
\end{aligned}
\end{equation}

Finally it is possible to write the above integrals in terms of conserving currents using the notations $\dt_\mu = (\dt/\dt T, \dt/\dt \s)$
\begin{equation}
 \label{7.1}
 \dot{I}_{m}=0 \rightarrow \dt_\m {j_m}^\m=0.
\end{equation}
hence, the currents read
\begin{equation}
 \label{7.2}
\begin{aligned}
   {j_{n}}^0&=\sum_{k=1}^n\a^k\b^{n-2k}t_{n,k},\\
   {j_{n}}^1&=-2\sum_{k=1}^n\a^k\b^{n-2k+1}\fr{n-k}{n-2k+1}t_{n,k},
\end{aligned}
\end{equation}
with $\a=\P J'$ and $\b=\P+J'$. According to the Noether theorem each conserving current is associated to a global symmetry of the system.

\section{Conclusion}

The main result present in this letter is the analysis of integrability properties of the corresponding Hamiltonian system and observation that these are related to Motzkin paths.

We start from a Hamiltonian of the form
\begin{equation}
H=\int_{-\pi}^{\pi}d\s  \, \left[\P^2J'+\P{J'}^2\right],
\end{equation}
which is motivated by studies of the exact renormalization group flows of $N\times N$ matrix scalar field theories \cite{Akhmedov:2010mz}. In particular, in the limit $J \ll \P$ (together with all its derivatives) equations of motion for the above system become just
\begin{equation}
\dt_T\r= \r \dt_s \r, \quad \r= \fr{\dt_T{J}}{\dt_\s J},
\end{equation}
that is the known Burgess-Hopf equation describing dynamics of shock waves. On the field theory side this corresponds to taking IR limit and large $N$ limit as it has been shown in \cite{Akhmedov:2010mz}. So far there has not been found a direct path from renormalization group equations to the Hamiltonian in question, hence we consider it as a motivated toy-model.

Starting with this setup we show, that given the fundamental fields $J(\s,T)$  and their conjugates $\P(\s,T)$, this system has infinite number of conserved quantities $I_n$  of the form
\begin{equation}
\begin{aligned}
I_n&=\int_\s\sum_{k=1}^n\left(\P J'\right)^{k}\left(\P+J'\right)^{n-2k}T_{n-2,k-1}+\left(\P+J'\right) \d_{n}^1,
\end{aligned}
\end{equation}
which Poisson-commute with each other. This allows to conjecture, that the non-linear system in question might be indeed integrable. A way to show that explicitly would be to find the corresponding Lax pair or to satisfy the Painlev\'e criterion. 

What is more interesting about the presented result is the appearance of the so-called Motzkin number $T_{n,k}$ (sequence A055151 in \cite{Sloane}) in the integrals of motion $I_n$. Initially, these numbers and the corresponding polynomials appear in the problem of counting all routes on a lattice with a given number of vertical and horizontal steps (see Figure \ref{Motz}). Hence, each integral of motion $I_n$ is given by a Motzkin polynomial corresponding to a path of length $n-2$ with $k-1$ horizontal steps.

An interesting problem would be to consider other polynomials corresponding to various paths on a lattice, and reversely build a set of expressions understood as integrals of motion for some system. Having such procedure would be a fascinating way of generating dynamical systems. If it becomes possible to strictly prove classical integrability of the system, one may speculate on whether having description of the corresponding integrals of motion in terms of some polynomials is equivalent to integrability. 

On the other hand the above logic may be reverted, and one may ask whether integrals of motion of a given Hamiltonian system correspond to some polynomials not known before. This conjecture can be tested explicitly on some simple examples. 

Finally, on the field theoretical side one is still interested in finding a way to show that the Motzkin Hamiltonian indeed describes exact renormalization group flow of a theory, probably under some assumptions.

\section*{Acknowledegments}

The work of ETM was supported by the Alexander von Humboldt Foundation and in part by the Russian Government programme of competitive growth of Kazan Federal University. 

ETM would like to thank CEA/Saclay, where some part of the work was done, and personally Mariana Gra\~na  for warm hospitality and nice working atmosphere. Authors are grateful to Emil Akhmedov for
enlightening discussions on the subject.

\appendix

\section{Derivation of integrals of motion}
\label{deriv1}

Let us start by listing few first integrals of motion for which purpose we first introduce the notations
\begin{equation}
 \label{a2}
 \begin{split}
  & \a=\P J';\\
  & \b=\P+J'.
 \end{split}
\end{equation}
The equations of motion for these variables are rather simple
\begin{equation}
 \label{a3}
\begin{split}
 & \dot{\a}=2(\a\b)';\\
 & \dot{\b}=\left(\b^2+2\a\right)',
\end{split}
\end{equation}
using which one easily checks that the following expressions represent conserved charges
\begin{equation}
\label{exiom}
\begin{aligned}
I_1&=\int_\s\b,\\
I_2&=\int_\s\a,\\
I_3&=\int_\s\a\b,\\
I_4&=\int_\s\a^2+\a\b,\\
I_5&=\int_\s3\a^2\b+\a\b^3,\\
I_6&=\int_\s2\a^3+6\a^2\b^2+\a\b.
\end{aligned}
\end{equation}
One immediately notices that all terms in each expression are of the same power in the fields $\P$ and $J'$. After observing some other patters above one conjectures the following general expressions for an integral of motion
\begin{equation}
 \label{a4}
\begin{aligned}
I_{n}&=\int_\s\sum_{k=1}^{n}\a^k\b^{n-k}t_{n,k},
\end{aligned}
\end{equation}
where the coefficients $t_{n,k}$ are constrained to satisfy certain conditions. 

Indeed, let us show that these expressions represent an infinite number of conserving charges and find the coefficients explicitly. Hence, we consider time derivative
\begin{equation}
\label{dot}
\begin{aligned}
\dot{I}_{n}=&\int_\s\sum_{k=1}^n2k\a^{k-1}\a'\b^{n-2k+1}t_{n,k}+\sum_{k=2}^{n+1}2(n-2k+2)\a^{k-1}\b^{n-2k+1}\a't_{n,k-1}\\
&+\sum_{k=1}^n2(n-2k)\a^k\b^{n-2k}\b' t_{n,k}\\
=&\int_\s \sum_{k=1}^n\big(2k t_{n,k}+2(n-2k+2)t_{n,k-1}\big)\a^{k-1}\a'\b^{n-2k+1}-2n\b^{n-1}\a't_{n,0}\\
&+\sum_{k=1}^n2(n-k)\a^k\b^{n-2k}\b' t_{n,k}-2n \a^n \a' \b^{-n-1}t_{n,n}
\end{aligned}
\end{equation}
Here in the first line we used the equations of motion for $\a$ and $\b$ and shifted the summation index in the second term, while in the second line we added and subtracted the term with $k=1$ needed to complete the second sum. Now to form a full derivative and to make the additional terms vanish one imposes the following conditions for the coefficients 
\begin{equation}
\label{t}
\begin{aligned}
 \fr{k-1}{n-2k+1}t_{n,k}&=\fr{n-2k+2}{k}t_{n,k-1},\\
 t_{n,0}&=0,\\
 t_{n,n}&=0
\end{aligned}
\end{equation}
The second condition above ensures that all terms in the charges $I_{n}$ always have at least one power of $\a$ as it can be explicitly seen from \eqref{exiom} while the last condition removes terms of negative powers from \eqref{dot}. In what follows this will be extended to \{$t_{n,k}=0$ for all $2k > n$ \}.

The recurrence relations \eqref{t} can be used to determine explicit expressions for the coefficients $t_{n,k}$ as follows
\begin{equation}
\begin{aligned}
t_{n,k}&=\fr{(n-2k+1)(n-2k+2)}{k(k-1)}t_{n,k-1}\\
&=\fr{(n-2k+1)(n-2k+2)(n-2k+3)(n-2k+4)\cdots (n-3)(n-2)}{k(k-1)(k-1)(k-2)\cdots 2\cdot1}t_{n,1}\\
&=\fr{(n-2)!}{(n-2k)!k!(k-1)!}t_{n,1}.
\end{aligned}
\end{equation}
Finally, using the freedom to choose the overall normalization of each of $I_{n}$ one is allowed to fix $t_{n,1}=1$ 
\begin{equation}
\begin{aligned}
t_{n,k}&=\fr{(n-2)!}{(n-2k)!k!(k-1)!}.
\end{aligned}
\end{equation}
Since this formula does not cover the case $n=1$ and the cases $2k > n$ we set the additional constraints based on the explicit form of the integrals of motion \eqref{exiom}
\begin{equation}
\begin{aligned}
t_{1,1}&=1,\\
t_{n,k}&=0, \forall k>2n.
\end{aligned}
\end{equation}

\section{Poisson brackets}

For proper integrable systems one observes all integrals of motion in involution meaning that Poisson brackets $\{I_n,I_m\}$ vanishes for any $I_n$ and $I_m$. Since the fundamental variables for our theory are $\P(T,\s)$ and $J(T,\s)$ the Poisson bracket is written as
\begin{equation}
\{F,G\}=\int_\s \fr{\d F}{\d J(T,\s)}\fr{\d G}{\d \P(T,\s)}-\fr{\d G}{\d J(T,\s)}\fr{\d F}{\d \P(T,\s)},
\end{equation}
where $F$ and $G$ are some functionals in $\P$ and $J$. In what follows we will not mention dependence on $T$ and $\s$ for the sake of space.

Let us now show that $\{I_n,I_m\}=0$ for any $n,m$, and start with derivatives of $I_n$ with respect to the fundamental variables. Denoting variation with respect to a function $f$ by $\d_f$ we note the  following
\begin{equation}
\begin{split}
\d_J I_m&=-\dt_\s \d_{J'}I_m,\\
\d_{J'}I_m&=\d_\a I_m \P+\d_\b I_m,\\
\d_{\P}I_m&=\d_\a I_m J'+\d_\b I_m.
\end{split}
\end{equation}
Given these we can write for the Poisson bracket
\begin{equation}
\begin{aligned}
\{I_m,I_n\}&=2\int_\s\d_J I_{[m} \d_\P I_{n]}=-2\int_\s\dt_\s\big(\d_{J'} I_{[m}\big)\d_\P I_{n]}\\
&=-2\int_\s \dt_\s \big(\d_\a I_m \P+\d_\b I_m\big)\big(\d_\a I_n J'+\d_\b I_n\big)\\
&=-2\int_\s \a(\d_\a I_m)'\d_\a I_n+\P'\d_\a I_m \d_\b I_n +J'(\d_\b I_m)'\d_\a I_n\\
&+(\d_\b I_m)'\d_\b I_n+\P(\d_\a I_m)'\d_\b I_n\\
&=-2\int_\s\Big[\a (\d_\a I_m)'\d_\a I_n+\b'\d_\a I_m \d_\b I_n +\b (\d_\a I_m)'\d_\b I_n+(\d_\b I_m)'\d_\b I_n\Big],
\end{aligned}
\end{equation}
where antisymmetrization in $\{m,n\}$ is always undermined. Here we used integration by parts and the antisymmetry properties to recollect terms with $J,\P$ and their derivatives back into $\a$ and $\b$.

Substituting the explicit form of the integrals $I_m$ and introducing a new variable $\g=\a \b^{-2}$ for convenience the integrand of the above expression can be written as (all terms are antisymmetric in $\{m,n\}$)
\begin{equation}
\begin{aligned}
\sum_{k,l}t_{n,k}t_{m,l}&\Big[kl\g^k\b^n(\g^{l-1}\b^{m-2})'+(m-2l)k \g^{k-1}\b^{n-1}(\g^l \b^{m-1})'\\
&+(n-2k)(m-2l)(\g^l \b^{m-1})'\g^k \b^{n-1}\Big]-(n\leftrightarrow m)\\
=\sum_{k,l}t_{n,k}t_{m,l}&\Big[kl\g^{k-1}\b^{n-1}(\g^l\b^{m-1})'+(m-2l)k \g^{k-1}\b^{n-1}(\g^l \b^{m-1})'\\
&+(n-2k)(m-2l)(\g^l \b^{m-1})'\g^k \b^{n-1}\Big]-(n\leftrightarrow m)\\
=\sum_{k,l}t_{n,k}t_{m,l}&\Big[(m-l)(m-1)k \g^{k+l-1}+(n-2k)(m-2l)(m-1)\g^{k+l} \Big]\b^{n+m-3}\b'\\
+t_{n,k}t_{m,l}&\Big[(m-l)kl\g^{k+l-2}+l(n-2k)(m-2l)\g^{k+l-1}\Big]\g'\b^{n+m-2}-(n\leftrightarrow m)
\end{aligned}
\end{equation}
where in the second line we the antisymmetry to shift  powers of $\b$ and $\g$ out of the derivative in the first term. Noticing that the power of $\g$ in the second term in each line is just that of the first term shifted as $k \to k+1$ we can use the property of the Motzkin coefficients $k(k-1)t_{n,k}=(n-2k+1)(n-2k+2)t_{n,k-1}$ to write the above  expression as
\begin{equation}
\begin{aligned}
\{I_n,I_m\}&=\sum_{k,l}t_{n,k}t_{m,l}\Big[A_{n,m,k,l}\g^{k+l-2}\g'\b^{n+m-2}+B_{n,m,k,l}\g^{k+l-1}\b^{n+m-3}\b'\Big]-(n\leftrightarrow m),\\
\mbox{where}&\\
A_{n,m,k,l}&=\fr{kl(m-l)(n+2k+1)+kl(k-1)(m-2l)}{n-2k+1},\\
B_{n,m,k,l}&=\fr{k(m-l)(m-1)(n+2k+1)+k(k-1)(m-2l)(m-1)}{n-2k+1}.
\end{aligned}
\end{equation} 
This sum does not form a full derivative term by term and one must turn to summation over $p=k+l$ to actually get cancellation. Hence, we write
\begin{equation}
\label{inv_last}
\begin{aligned}
\{I_n,I_m\}&=\sum_{p=2}^N\sum_{k=1}^{p-1}\Big[\hat{A}_{n,N,k,p}\g^{p-2}\g'\b^{N-2}+\hat{B}_{n,N,k,p}\g^{p-1}\b^{N-3}\b'\Big],\\
\mbox{with}\\
\hat{A}_{n,N,k,p}&=A_{n,N-n,k,p-k}-A_{N-n,n,p-k,k},\\
\hat{B}_{n,N,k,p}&=B_{n,N-n,k,p-k}-B_{N-n,n,p-k,k},
\end{aligned}
\end{equation}
and $N=n+m$. Although each term in the sum has now the same power of the variables $\g$ and $\b$ the full derivative can be obtained only after taking the summation along $k$ explicitly. This is a tough calculational task and it is much easier to check
\begin{equation}
\sum_{k=1}^{p-1}\fr{\hat{A}_{n,N,k,p}}{p-1}-\sum_{k=1}^{p-1}\fr{\hat{B}_{n,N,k,p}}{N-2}=0.
\end{equation}
Indeed, using Wolfram Mathematica and performing the calculation explicitly one gets the desired cancellation for any $p,N$ and $n$. Obviously, this ensures that the expression \eqref{inv_last} is indeed a full derivative and hence all the integrals of motion are in involution.

\bibliography{bib}
\bibliographystyle{utphys}

\end{document}